\documentclass[apjl]{emulateapj}

\usepackage{graphicx}
\usepackage{t1enc}
\usepackage[varg]{txfonts}
\usepackage{epsfig}
\usepackage{rotating}
\usepackage{natbib}
\usepackage{color}

\newcommand{\mum}   {$\mu$m}
\newcommand{\kms}   {km~s$^{-1}$}

\newcommand{\jpb}   {$\rm Jy~beam^{-1}$}    
\newcommand{\lo}    {$L_{\sun}$}
\newcommand{\mo}    {$M_{\sun}$}

\newcommand{\et}    {et al.}
\newcommand{\eg}    {e.\,g.,}

\newcommand{\hii}   {H{\small II}}

\definecolor{RED}{rgb}{1.0,0.0,0.0}

\shorttitle{The intermediate-mass hot core IRAS~22198+6336}
\shortauthors{S\'anchez-Monge et al.}

\begin{document}

\title{IRAS~22198+6336: Discovery of an Intermediate-Mass Hot Core}

\author{\'Alvaro S\'anchez-Monge\altaffilmark{1},
        Aina Palau\altaffilmark{2},
	Robert Estalella\altaffilmark{1},
	Stan Kurtz\altaffilmark{3},
	Qizhou Zhang\altaffilmark{4},
	James Di Francesco\altaffilmark{5},
	and
	Debra Shepherd\altaffilmark{6}}
\altaffiltext{1}{Departament d'Astronomia i Meteorologia (IEEC-UB), Institut de
Ci\`encies del Cosmos, Universitat de Barcelona, Mart\'i i Franqu\`es, 1,
E-08028 Barcelona, Spain}
\email{asanchez@am.ub.es}
\altaffiltext{2}{Institut de Ci\`encies de l'Espai (CSIC-IEEC), Campus UAB -- Facultat de Ci\`encies, Torre C5 -- parell 2, E-08193 Bellaterra, Catalunya, Spain}
\altaffiltext{3}{Centro de Radioastronom\'ia y Astrof\'isica, Universidad
Nacional Aut\'onoma de M\'exico, Apdo. Postal 3-72, 58090, Morelia, Michoac\'an,
Mexico}
\altaffiltext{4}{Harvard-Smithsonian Center for Astrophysics, 60 Garden Street, Cambridge, MA 02138, USA}
\altaffiltext{5}{Conseil national de recherches Canada, Institut Herzberg d'astrophysique, 5071 West Saanich Road, Victoria BC, Canada}
\altaffiltext{6}{NRAO, PO Box O, Socorro, NM 87801-0387, USA}

\begin{abstract}
We present new SMA and PdBI observations of the intermediate-mass object
IRAS~22198+6336 in the millimeter continuum and in several molecular line
transitions. The millimeter continuum emission reveals a strong and compact
source with a mass of $\sim\!5$~\mo\ and with properties of Class~0 objects. CO
emission shows an outflow with a quadrupolar morphology centered on the position
of the dust condensation. The CO outflow emission seems to come from two
distinct outflows, one of them associated with SiO outflow emission. A large set
of molecular lines has been detected toward a compact dense core clearly
coincident with the compact millimeter source, and showing a velocity gradient
perpendicular to the outflow traced by CO and SiO. The chemically rich spectrum
and the rotational temperatures derived from CH$_3$CN and CH$_3$OH (100--150~K)
indicate that IRAS~22198+6336 is harbouring one the few intermediate-mass hot
cores known at present.
\end{abstract}


\section{Introduction \label{sint}}

Early stages of massive star formation have numerous observational signatures,
including molecular masers and outflows, sub-millimeter continuum sources, hot
molecular cores, and hyper/ultracompact \hii\ regions. Hot molecular cores
(HMCs), first discovered in the vicinity of massive protostars (see Kurtz \et\
2000 for a review), are compact ($\le\!0.1$~pc, $n\gtrsim\!10^7$~cm$^{-3}$)
objects with relatively high temperatures ($T_\mathrm{k}\gtrsim\!100$~K) that
show a very rich chemistry in complex organic molecules (CH$_3$CN, CH$_3$OH,
CH$_3$OCHO,\ldots).

Recently, regions characterized by high temperatures and densities have also
been detected around low-mass protostars (see Ceccarelli 2004 for a review).
These so-called hot corinos share many characteristics with HMCs, although the
former are $\sim\!10^2$ times less massive. Many questions remain unanswered
regarding both the high-mass and low-mass hot core phase. For example, HMCs are
relatively large objects ($\sim$ 0.1 pc) and harbor clusters of stars, while hot
corinos seem to be smaller molecular envelopes surrounding a single low-mass
young stellar object (YSO) or binary. Additionally, although HMCs are generally
understood to be illuminated by the newly born high-mass stars embedded within
the cores themselves (Cesaroni 2005), an alternative scenario has been proposed
for hot corinos, in which jets and shocks are responsible for enhancing the
temperature and the abundances of many species (Chandler \et\ 2005).

Intermediate-mass YSOs (IMYSOs; $M_\star\sim\!2$--8~\mo) are excellent targets
to study the different hot core heating mechanisms (radiative or shocks),
because they share some properties of massive stars, such as their association
with clusters and their ability to photodissociate and ionize the surrounding
gas, and present the advantages of being located closer ($\le1$~kpc) than most
of the massive star-forming regions, and in regions of less complexity. However,
there are only two IMYSOs with hot core emission reported in the literature,
NGC\,7129-FIRS~2 and IC~1396~N (Fuente \et\ 2005, 2009), and few conclusions can
be inferred from such a small number of intermediate-mass hot cores (IMHCs).

IRAS~22198+6336 (hereafter I22198) is an IMYSO located at a distance of 764~pc
(Hirota \et\ 2008) with a bolometric luminosity of 370~\lo. The presence of a
strong and compact submillimeter dust condensation (Jenness \et\ 1995) with no
near-infrared nor mid-infrared emission suggests that it is a deeply embedded
object, classified as an intermediate-mass Class~0 source by S\'anchez-Monge
\et\ (2008). Single-dish telescope observations reveal a CO outflow well
centered on the position of the dust condensation (Zhang \et\ 2005).
S\'anchez-Monge \et\ (2008) detect slightly resolved partially optically thick
centimeter emission consistent with a thermal ionized wind (or radiojet).
Additionally, NH$_3$ and CS dense gas emission together with H$_2$O and OH maser
emission has been detected toward I22198 (Tafalla \et\ 1993; Larionov \et\ 1999;
Valdettaro \et\ 2002; Edris \et\ 2007; Hirota \et\ 2008), all indicative of a
very early evolutionary stage.

In this letter, we present new interferometric continuum and molecular line
observations that reveal that I22198 is associated with an intermediate-mass hot
core.

\begin{figure*}[t!]
\begin{center}
\begin{tabular}[b]{c c}
	\epsfig{file=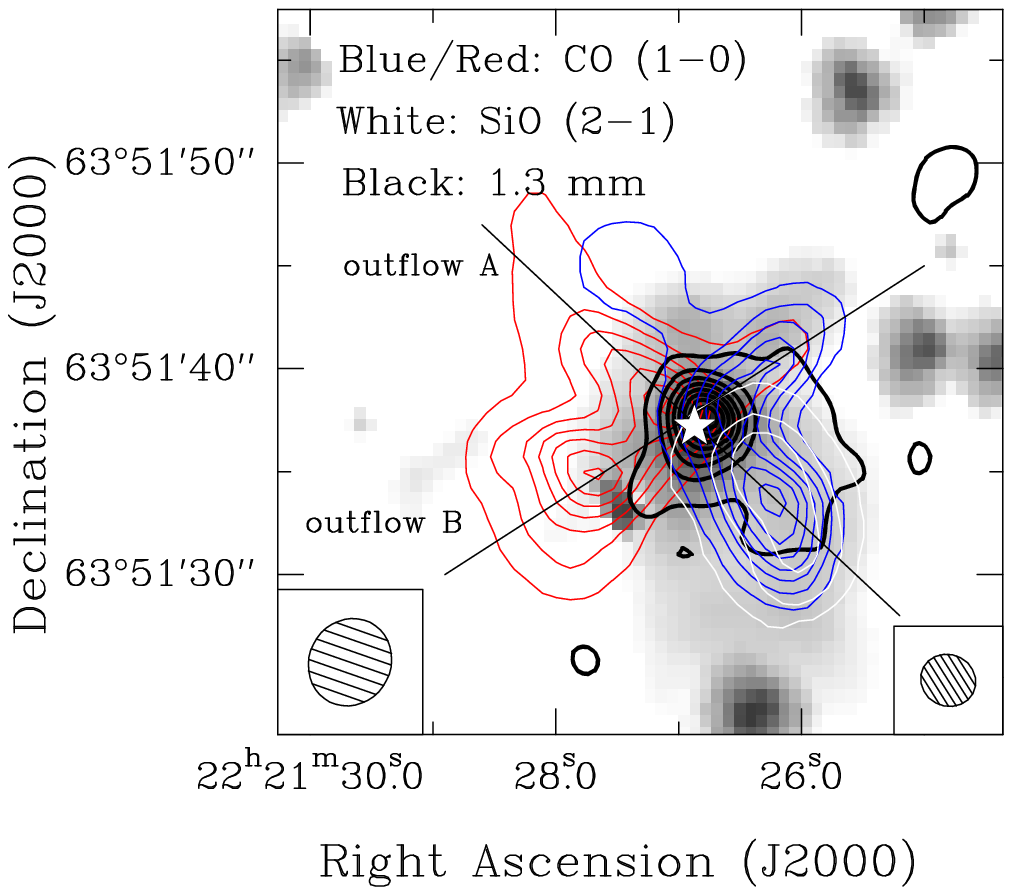, scale=0.7}	&
	\epsfig{file=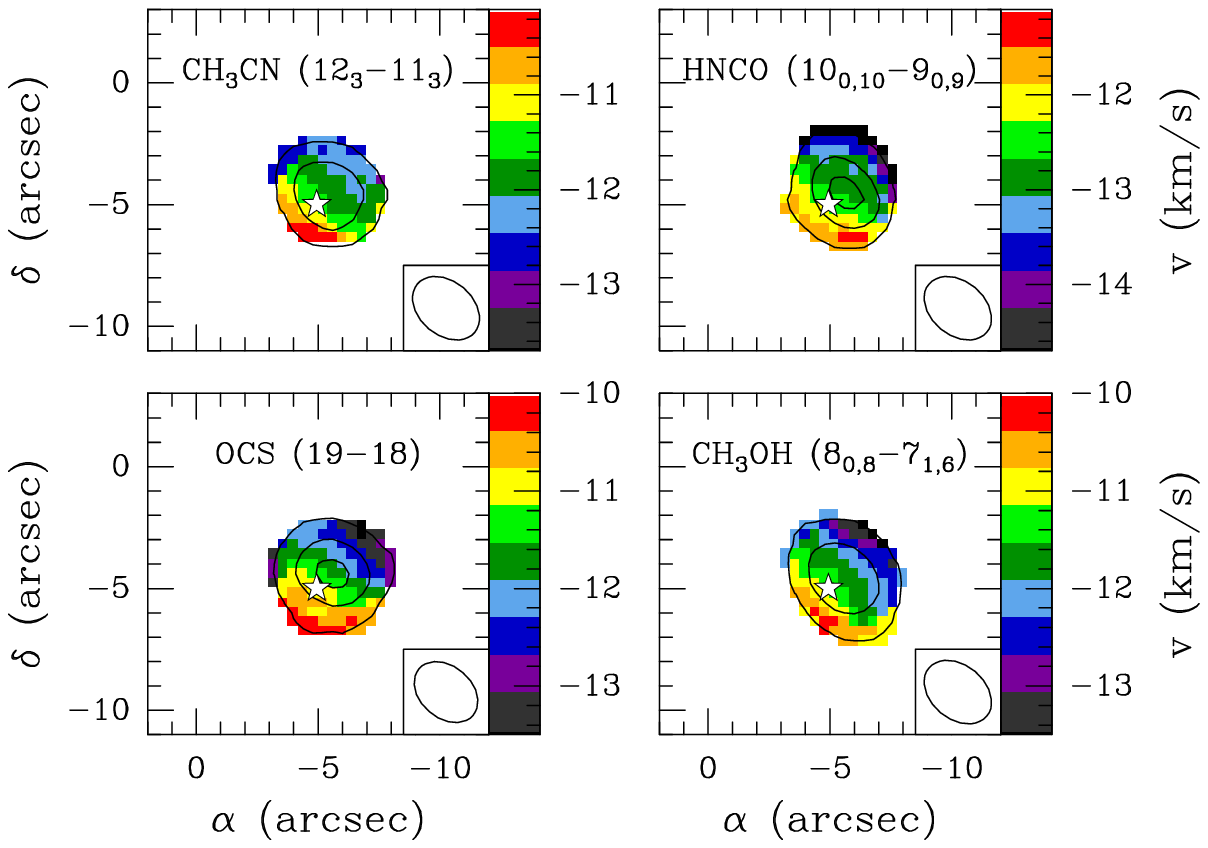, scale=0.7}	\\
\end{tabular}
\caption{IRAS~22198+6336.
  {\it Left}: Blue and red contours: PdBI CO\,(1--0) intensity map for
  blueshifted velocities (from $-5$ to $-15$~\kms\ with respect to the systemic
  velocity, $\varv_\mathrm{LSR}\simeq\!-11.3$~\kms) and redshifted velocities
  (from $+5$ to $+15$~\kms\ with respect to the $\varv_\mathrm{LSR}$). Levels
  start at 5\%, increasing in steps of 15\% of the peak temperature,
  7200~K~\kms\ and 5400~K~\kms\ for blue and red respectively. White contours:
  PdBI SiO\,(2--1) intensity map for blueshifted velocities (from $-5$ to
  $-15$~\kms\ with respect to the $\varv_\mathrm{LSR}$). Levels start at 5\%,
  increasing in steps of 30\% of the peak temperature, 450~K~\kms. Synthesized
  beams for CO and SiO images are shown in the bottom right and left corners,
  respectively. Black contours: SMA 1.3~mm continuum map. Levels are $-4$, and
  $4$ to $60$ in steps of 8, times 3.92~m\jpb. Grey scale: 4.5~\mum\
  IRAC/Spitzer image. {\it Right}: Each of the four panels show in black
  contours the zero-order moment (intensity) map, and in colour scale the
  first-order moment (velocity) map for four molecular transitions observed with
  the SMA. Levels are 5\%, 45\%, and 85\% the intensity peak 2.67, 3.20, 3.32,
  3.97~K~\kms, for CH$_3$CN, HNCO, OCS and CH$_3$OH respectively. The (0,0)
  position corresponds to the SMA phase center. The white star marks the
  position of the centimeter continuum source detected by S\'anchez-Monge \et\
  (2008).}
\label{firegion}
\end{center}
\end{figure*}

\section{Observations \label{sobs}}

We observed the I22198 region with the Submillimeter Array (SMA\footnote{The SMA
is a joint project between the Smithsonian Astrophysical Observatory and the
Academia Sinica Institute of Astronomy and Astrophysics, and is funded by the
Smithsonian Institution and the Academia Sinica.}; Ho \et\ 2004) in the 1.3~mm
(230~GHz) band using the compact array configuration on 2008 June 11. The phase
center was $\alpha\rm{(J2000)}=22^{\mathrm h}21^{\mathrm m}27\fs617$, and
$\delta\rm{(J2000)}=+63\degr51\arcmin42\farcs18$, and the projected baselines
ranged from 7~k$\lambda$ to 100~k$\lambda$. The two sidebands of the SMA covered
the frequency ranges of 219.4--221.4~GHz and 229.4--231.4~GHz, with a spectral
resolution of $\sim\!0.5$~\kms. System temperatures ranged between 200 and
300~K. The zenith opacities at 225~GHz were around 0.15 and 0.20 during the
track. The FWHM of the primary beam at 1.3~mm was $\sim\!56$\arcsec. Bandpass
calibration was performed by observing the quasar 3C454.3. Amplitude and phase
calibrations were achieved by monitoring 0019+734 and 1928+739, resulting in an
rms phase of $\sim\!30$\degr. The absolute flux density scale was determined
from Callisto and Jupiter with an estimated uncertainty around 15\%. Data were
calibrated and imaged with the MIRIAD software package. The continuum was
constructed in the ({\it u,v}) domain from the line-free channels. Imaging was
performed using natural weighting, obtaining a synthesized beam of
$3\farcs0\times2\farcs2$ with a P.A.=50\degr, and 1~$\sigma$ rms of
80~mJy~beam$^{-1}$ per channel, and 3.9~mJy~beam$^{-1}$ for the continuum.

The IRAM Plateau de Bure Interferometer (PdBI\footnote{The Plateau de Bure
Interferometer (PdBI) is operated by the Institut de Radioastronomie
Millimetrique (IRAM), which is supported by INSU/CNRS (France), MPG (Germany),
and IGN (Spain).}) was used to observe the CO\,(1--0) (115.12~GHz) and
SiO\,(2--1) (86.85~GHz) molecular transitions toward I22198. We carried out
2-pointing mosaic observations (primary beam$\simeq\!56$\arcsec), in 2008 and
2009, with the array in the C and D configurations. The two spectral setups,
tuned at 115.27~GHz (2.7~mm) and at 86.85~GHz (3.5~mm), include several
molecular transitions, which will be presented in a forthcoming paper
(S\'anchez-Monge \et, in prep.). For the CO and SiO we used a correlator unit of
40~MHz of bandwidth with 512 spectral channels, which provides a spectral
resolution of $\sim\!0.3$~\kms. A number of 320~MHz continuum units were used to
image the continuum, six at 115~GHz and three at 86~GHz. The typical system
temperatures for the receivers were $\sim\!200$~K at 115~GHz and $\sim\!100$~K
at 86~GHz. Bandpass calibration was performed by observing 3C273 and 3C454.3.
Amplitude and phase calibrations were achieved by monitoring 0116+731, 1928+738
and 2037+511, resulting in a phase rms of around $25$\degr. The absolute flux
density scale was determined from MWC349 and 3C84, with an estimated uncertainty
around 25\%. The data were calibrated with the program CLIC, and imaged with
MAPPING, both parts of the GILDAS\footnote{GILDAS: Grenoble Image and Line Data
Analysis System, see http://www.iram.fr/IRAMFR/GILDAS.} software package.
Imaging was performed using natural weighting, obtaining a synthesized beam of
$3\farcs0\times2\farcs5$, with a P.A.=107\degr\ at 115~GHz, and
$4\farcs5\times4\farcs0$, with a P.A.=81\degr\ at 87~GHz. The 1~$\sigma$ rms
for the channel and continuum images were $40$ and $0.6$~m\jpb\ at 115~GHz, and
$30$ and $0.3$~m\jpb\ at 87~GHz, respectivelly.

\section{Results \label{sres}}

\subsection{Continuum results \label{srescont}}

In Figure~\ref{firegion} (left panel) we show the interferometric millimeter
continuum map toward I22198 (black contours). We detected a compact source
(with coordinates $\alpha=22^{\mathrm h}21^{\mathrm m}26\fs78$ and
$\delta=+63\degr51\arcmin37\farcs5$) coincident with the centimeter source
reported by S\'anchez-Monge \et\ (2008), and surrounded by a faint structure
extended toward the southwest. The source is clearly detected at 3.5, 2.7, and
1.3~mm with integrated flux densities of $24\pm6$~mJy, $55\pm15$~mJy, and
$500\pm75$~mJy, respectively, and has a similar morphology at the three
wavelengths. The deconvolved size of the source is $3\farcs0\times2\farcs1$,
with P.A.=$150$\degr, corresponding to $2300\times1600$~AU. The millimeter
continuum emission is likely tracing the dust envelope of I22198.

\begin{figure}[t!]
\begin{center}
\begin{tabular}[b]{c}
	\epsfig{file=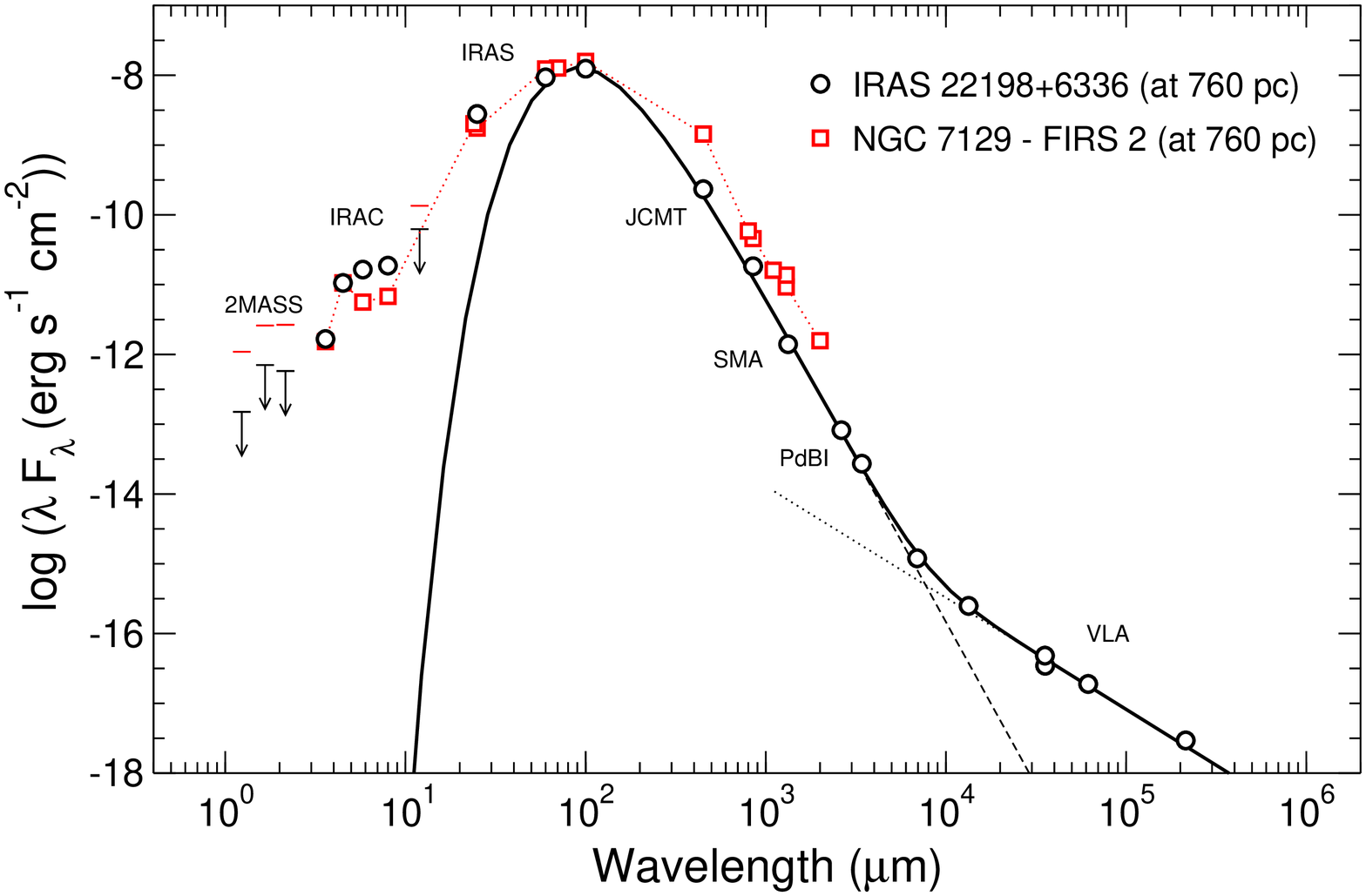, scale=0.3}	\\
\end{tabular}
\caption{Spectral energy distribution of IRAS~22198+6336 (black circles). Solid
  line: sum of the modified blackbody law and the thermal ionized wind. Black
  dashed line: modified blackbody law for the dust envelope with a dust
  emissivity index of $\beta=1.7$, source radius of $3\farcs0$, dust temperature
  of 36~K, dust mass of 5~\mo, and a dust mass opacity coefficient of
  0.9~cm$^2$~g$^{-1}$ at 1.3~mm. Black dotted line: thermal free-free emision
  from an ionized wind with a spectral index of $\sim\!0.6$ at centimeter
  wavelengths. We estimated the total bolometric luminosity, from IR up to mm
  wavelengths, to be $L_\mathrm{bol}\simeq370$~\lo\ for a distance of 0.76~kpc.
  Red squares: NGC~7129-FIRS~2 data (from Eiroa \et\ 1998, Crimier \et\ 2010,
  and 2MASS and IRAC catalogues) rescaled at the distance of I22198 (760~pc). We
  estimated $L_\mathrm{bol}\simeq480$~\lo. Labels of telescopes refer only to
  I22198 data.}
\label{fised}
\end{center}
\end{figure}

The spectral energy distribution (SED) reported by S\'anchez-Monge \et\ (2008)
has been improved with the new continuum data. We also included the  fluxes
between 3.6~\mum\ and 8.0~\mum\ extracted from the Spitzer Space Telescope
Infrared Array Camera (IRAC) maps (PI: Giovanni Fazio, project: 40147). We
retrieved the images using the
Leopard\footnote{http://ssc.spitzer.caltech.edu/warmmission/propkit/spot/}
software, and evaluated the fluxes using the
MOPEX\footnote{http://ssc.spitzer.caltech.edu/dataanalysistools/tools/mopex/}
software. The global SED (open circles in Fig.~\ref{fised}) was fitted by a
modified blackbody law plus a thermal ionized wind (with a spectral index of
$\sim\!0.6$ at centimeter wavelengths). Assuming a dust mass opacity coefficient
given by $\kappa_\nu=\kappa_0(\nu/\nu_0)^\beta$ with
$\kappa_0=0.9$~cm$^{2}$~g$^{-1}$ at $\nu_0=230$~GHz (Ossenkopf \& Henning 1994),
the SED is well fitted with a dust emissivity index of $1.7\pm0.2$, a dust
temperature of $36\pm2$~K, and an envelope mass of $5\pm1$~\mo. As shown in
Fig.~\ref{fised}, the SED of I22198 resembles that of Class~0 objects (Andr\'e
\et\ 1993).

\subsection{Molecular results \label{sresmol}}

Interferometric $^{12}$CO\,(1--0) (Fig.~\ref{firegion}~left) and
$^{12}$CO\,(2--1) (S\'anchez-Monge \et, in prep.) maps reveal an outflow with a
quadrupolar morphology clearly centered on the position of the dust
condensation. The outflow emission spans a velocity range from $-25$~\kms\ to
$+5$~\kms\ for the $J=1\to0$ transition (the systemic velocity is
$\varv_\mathrm{LSR}\simeq\!-11.3$~\kms). The quadrupolar morphology of the
outflow can be interpreted as the walls of a single outflow or as the
superposition of two bipolar outflows: outflow A in the southwest-northeast
direction, and outflow B in the northwest-southeast direction. Both outflows are
centered on the position of the millimeter continuum source, which could be a
millimeter clump harbouring at least two sources. In fact, higher angular
resolution observations (0.4'' or 300 AU) with the PdBI at 1.3~mm  reveal a
strong source with indications of a faint extension to the south (Palau \et, in
prep.). In the following, we will assume the two outflow interpretation. We
estimated the physical parameters of outflows A and B (see Table~\ref{timhcs})
and found values similar to the outflow parameters found in other IMYSOs (\eg\
IC~1396~N: Beltr\'an \et\ 2002; IRAS~22272+6358~A: Beltr\'an \et\ 2006). The
integrated SiO\,(2--1) emission, spanning a velocity range from $-25$ to
$-8$~\kms, is shown in white contours in Fig.~\ref{firegion}. This emission
coincides with the blueshifted lobe of outflow A. No redshifted counterpart is
observed for the SiO emission, however. The detection of SiO in outflow A
suggests that it is younger than outflow B, because gas-phase SiO abundances may
decrease with time (Klaassen \& Wilson 2007; Shang \et\ 2006). The SiO detection
also suggests that the CO quadrupolar morphology arises from two distinct
outflows, because SiO generally traces material associated with the primary jet
driving the outflow rather than the walls of the outflow cavity (\eg\ H211:
Palau \et\ 2006; Lee \et\ 2007). We also note that the IRAC 4.5~\mum\ band (grey
scale in Fig.~\ref{firegion}~left) shows extended emission (``green fuzzy'')
associated mainly with the blueshifted CO outflow lobes. Such 4.5~\mum\ extended
emission is thought to be associated with shocked H$_2$ emission (De Buizer \&
Vacca 2010).

\begin{figure}[t!]
\begin{center}
\begin{tabular}[b]{c}
	\epsfig{file=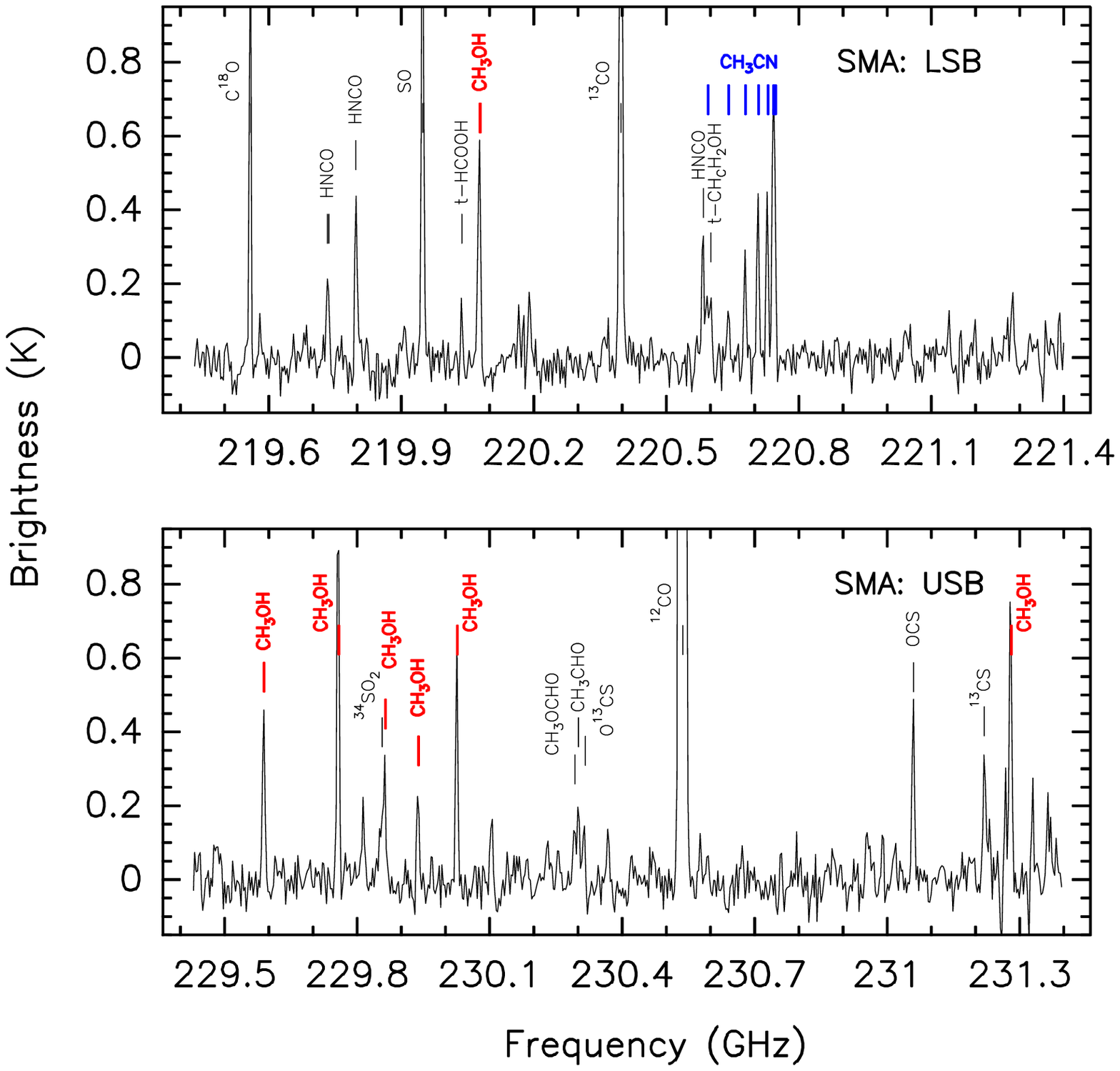, scale=0.5, angle=0}	\\
\end{tabular}
\caption{230~GHz continuum-free spectra in the image domain from the SMA
  data, for the lower side band (LSB) in the top panel and the upper side band
  (USB) in the bottom panel. The molecular transitions used in the rotational
  diagrams (Fig.~\ref{firotdia}) are shown in blue and red.}
\label{fispec}
\end{center}
\end{figure}

The wide-band SMA spectrum toward I22198 (Fig.~\ref{fispec}) reveals a
chemically rich dense core with typical hot core features (\eg\ Cesaroni \et\
1999; Gibb \et\ 2000). Molecular line transitions detected with
$T_\mathrm{B}\gtrsim0.2$~K have been identified and imaged. We note that the
linewidths observed for most of the molecules are 5--8~\kms.
Figure~\ref{firegion} (right panel) shows the maps of high-density tracers with
a compact source ($2\farcs1\times1\farcs3$, with $\mathrm{P.A.}=15$\degr) at the
position of the millimeter continuum source. Although the maps of each molecule
show a source barely resolved, and higher angular resolutions should be made,
first-order moment (velocity field) maps for these high-density tracers
(Fig.~\ref{firegion}~right) show velocity gradients in the northwest-southeast
direction, perpendicular to outflow A, presumable the youngest outflow. Most of
these molecules trace disks and/or toroids in other massive YSOs (\eg\ Beltr\'an
\et\ 2006). However, the presence of outflow B perpendicular to outflow A makes
the interpretation of these velocity gradients uncertain. Assuming that these
gradients trace rotation, the dynamical mass has been calculated from the
expression
$M_\mathrm{dyn}=\varv_\mathrm{rot}^{2}~R_\mathrm{rot}/G~\mathrm{sin}^{2}i$,
where $\varv_\mathrm{rot}$ is the velocity estimated from the gradient,
$R_\mathrm{rot}$ the radius of the core, $G$ the gravitational constant, and $i$
the inclination angle assumed to be 90\degr\ (edge-on). In our case, with 
$\varv_\mathrm{rot}\simeq\!2$~\kms\ and $R_\mathrm{rot}\simeq\!1$~\arcsec\ for
the four molecules shown in Fig.~\ref{firegion}~right, the dynamical mass is
$\sim\!3.5$~\mo. Additionally, following the models of Palla \& Stahler (1993),
and assuming a luminosity for the object of $\sim\!370$~\lo\ (Fig.~\ref{fised}),
the mass of the embedded YSO is about 4--5~\mo. Thus, the mass of the internal
YSO estimated from both methods is $\sim\!4$~\mo, which is similar to the dust
envelope mass ($\sim\!5$~\mo), as is characterisc of Class~0 objects.

\begin{figure}[t!]
\begin{center}
\begin{tabular}[b]{c}
	\epsfig{file=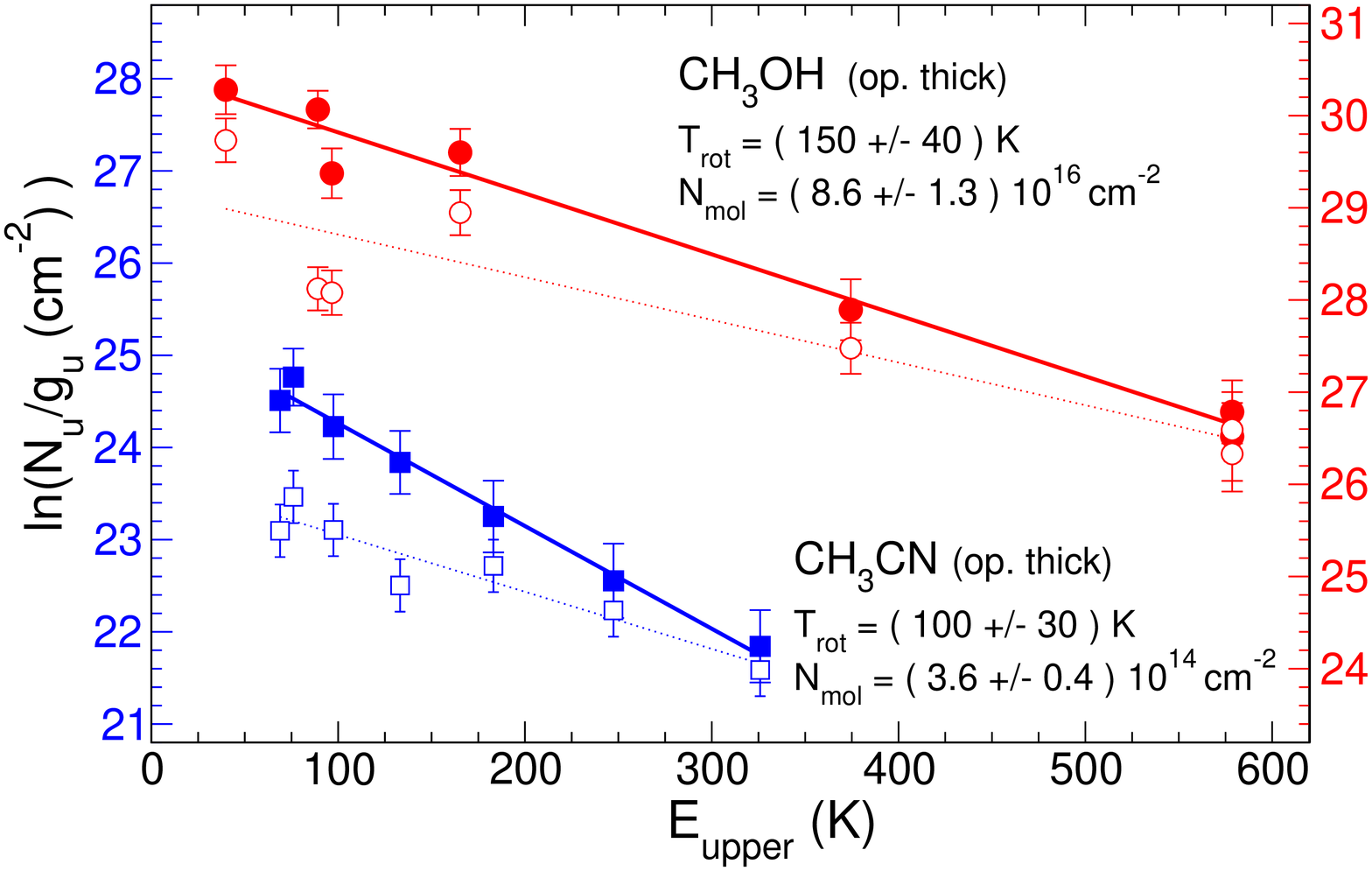, scale=0.30, angle=0}	\\
\end{tabular}
\caption{CH$_3$CN (blue squares) and CH$_3$OH (red circles) rotational diagrams.
  Open symbols and dashed lines refer to the optically thin approximation, while
  filled symbols and solid lines refer to the opacity which yields the  best
  linear fit. For the optically thick fit we give the values in the figure, and
  adopt them in the main text. For the optically thin fit,
  $T_\mathrm{rot}=160\pm30$~K and
  $N_\mathrm{mol}=(1.7\pm0.3)\times10^{14}$~cm$^{-2}$ for CH$_3$CN, and
  $T_\mathrm{rot}=220\pm60$~K and
  $N_\mathrm{mol}=(4.6\pm1.6)\times10^{16}$~cm$^{-2}$ for CH$_3$OH. Vertical
  left and right axes refer to CH$_3$CN and CH$_3$OH, respectively.}
\label{firotdia}
\end{center}
\end{figure}

Because of the wide SMA band, one can observe several CH$_3$CN and CH$_3$OH
transitions simultaneously. Following the rotational diagram method (Goldsmith
\& Langer 1999; Araya \et\ 2005), which assumes that all molecular levels are
populated according to the same excitation temperature, we can derive this
temperature, $T_\mathrm{rot}$, and the total column density, $N_\mathrm{mol}$,
for these molecules. We constructed the rotational diagrams following two
methods. First, we assumed optically thin emission (open symbols in
Fig.~\ref{firotdia}). Second, we estimated the opacity for each transition
(following Goicoechea \et\ 2006; see also Girart \et\ 2002) and we adopted the
value that yielded the best linear fit (filled symbols in Fig.~\ref{firotdia}).
The resulting opacities for the different CH$_3$CN and CH$_3$OH transitions
range from 0.3 to 3.5 and from 0.2 to 6.5, respectively. In the discussion we
will use the results of the latter fit, which are $T_\mathrm{rot}=100\pm30$~K
and $N_\mathrm{mol}=(3.6\pm0.4)\times10^{14}$~cm$^{-2}$ for CH$_3$CN, and
$T_\mathrm{rot}=150\pm40$~K and
$N_\mathrm{mol}=(8.6\pm1.3)\times10^{16}$~cm$^{-2}$ for CH$_3$OH. The high
temperatures, together with the chemically rich spectrum, are clear evidence
that I22198 is an intermediate-mass hot core.

\section{Discussion and Conclusions}

Our interferometric continuum and molecular line observations toward I22198
reveal a compact and strong millimeter source driving a quadrupolar outflow,
which is associated with a compact dense core with hot core signatures. At
infrared wavelengths, IRAC images reveal a single source coincident with the hot
core and with ``green fuzzies'' at 4.5~\mum. The high rotational temperature
(100--150~K) estimated from the CH$_3$CN and CH$_3$OH rotational diagrams, and
the chemically rich spectrum, make I22198 one of the few IMHCs known at present.
There are only two other IMHCs reported in the literature: NGC\,7129-FIRS~2
(Fuente \et\ 2005), and IC~1396~N (Neri \et\ 2007; Fuente \et\ 2009).
Table~\ref{timhcs} lists the main properties of the three IMHCs known to date.

IC~1396~N is associated with a single IRAS source, IRAS~21319+5802. However,
high-angular resolution millimeter images reveal a cluster of sources, only one
of which is associated with the IMHC (Neri \et\ 2007).  Moreover, while
Spitzer/IRAC observations reveal a counterpart for millimeter source 41.73+12.8,
there is no counterpart for the IMHC millimeter source (41.86+11.9; Choudhury
\et\ 2010; Neri \et\ 2007). All this makes it difficult to assess the IMHC
contribution to the IRAS and IRAC fluxes, and hence we refrain from comparing
IC~1396~N with I22198.

\begin{table}[t!]
\caption{Properties of intermediate-mass hot cores$^\mathrm{a}$}
\centering
\begin{tabular}{l c c c}
\hline\hline\noalign{\smallskip}	
					&IRAS~22198+6336		&NGC\,7129-FIRS~2	&IC~1396~N		\\
\hline
\noalign{\smallskip}
distance (pc)				&760				&1250			&750			\\
$L_\mathrm{bol}$ (\lo)			&370				&480			&<300			\\
$M_\mathrm{env}$ (\mo)			&5				&2			&<5			\\
clustering (mm)				&no				&no			&yes			\\
IRAC source				&yes				&yes			&no			\\
$L_\mathrm{cm}$ (mJy~kpc$^2$)		&0.34				&<0.36			&<0.15			\\
outflow					&yes				&yes			&yes			\\
\multicolumn{3}{l}{\phn\phn \it hot core information}	\\
$T_\mathrm{rot}$ (K)			&100--150			&50--100		&100			\\
$N_\mathrm{H_2}$ (cm$^{-2}$)		&$4\times10^{23}$~$^\mathrm{b}$	&$8\times10^{24}$	&$3\times10^{24}$	\\
X$_\mathrm{CH_3CN}$			&$1\times10^{-9}$		&$7\times10^{-9}$	&$5\times10^{-10}$	\\
\multicolumn{3}{l}{\phn\phn \it outflow information}	\\
$t_\mathrm{dyn}$ (yr)			&1300 / 1200			&4000			&2600			\\
$\dot{M}$ ($10^{-6}$\mo~yr$^{-1}$)	&9 / 8				&7			&50			\\
$P$ (\mo~\kms)				&0.3 / 0.2			&0.4			&4			\\
\hline
\end{tabular}
\begin{list}{}{}
\item[$^\mathrm{a}$] References are:
  I22198: S\'anchez-Monge \et\ 2008; this work (the two values in outflow
  parameters correspond to outflows A and B, respectively);
  NGC\,7129-FIRS~2: Fuente \et\ 2001, 2005, Crimier \et\ 2010 (centimeter data
  from the VLA project AR304, and outflow parameteres: S\'anchez-Monge,
  priv.~com.);
  IC~1396~N: Beltr\'an \et\ 2002; Neri \et\ 2007; Fuente \et\ 2009.
  Upper limits for IC~1396~N are due to the cluster properties (see text).
\item[$^\mathrm{b}$] Estimated following Fuente \et\ (2005), assuming $T=100$~K
  and a size of $2\farcs1\times1\farcs3$, which is the deconvolved size of the
  CH$_3$CN core.
\end{list}
\label{timhcs}
\end{table}

The other IMHC, NGC\,7129-FIRS~2, has a bolometric luminosity similar to I22198,
has a single associated source dominating the millimeter emission (Fuente \et\
2005) and also is clearly associated with a source detected in IRAC and MIPS
bands (S\'anchez-Monge, priv.~com.). Eiroa \et\ (1998) study the far-infrared
emission, and find one HIRES source, which is coincident with the MIPS source
(Crimier \et\ 2010). The association of the millimeter, far-infrared and
mid-infrared source with the IMHC is clear, as in the case of I22198 and,
similarly, ``green fuzzies'' in the 4.5~\mum\ IRAC image appear associated with
the molecular outflow, which is also quadrupolar (Fuente \et\ 2001) and has
physical parameters similar to the outflows in I22198 (Table~\ref{timhcs}).
Regarding the centimeter emission, the upper limit for NGC\,7129-FIRS~2 is
consistent with the centimeter emission found toward I22198
(Table~\ref{timhcs}). Furthermore, Figure~\ref{fised} shows that both IMHCs have
quite similar Class~0-like SEDs. Thus, NGC\,7129-FIRS~2 and I22198 have very
similar properties concerning clustering, centimeter and outflow emission, and
have a similar SED. Moreover, the hot core temperature, $T_\mathrm{rot}$, and
CH$_3$CN abundance, X$_\mathrm{CH3CN}$, are similar for these two IMHCs. At this
point, our data do not allow us to distinguish between the proposed hot core
scenarios (radiative or shocked), because the two IMHCs  have similar
luminosities and outflow/centimeter properties. A detailed study  of different
molecular abundances (enhanced by the radiation field or by shocks) will be
presented in a forthcoming paper, and should clarify the formation scenario of
the hot core.

In summary, the properties of the newly discovered IMHC, I22198, and its
comparison with the other known IMHCs help us better establish the
characteristics of the hot core phase for IMYSOs. We have found that the IMHC
phase is co-eval with mass ejection (as shown by outflow, ``green fuzzies'', and
faint centimeter emission), presents Class~0-like SEDs, has  hot core
temperatures $\sim\!100$~K, and X$_\mathrm{CH_3CN}\simeq\!10^{-9}$. A detailed
comparison of I22198 and NGC\,7129-FIRS~2 and the discovery of new IMHCs will
provide fundamental tools to understand the hot core phase of star formation.

\acknowledgments
\begin{small}
We thank the anonymous referee for his/her useful comments. \'A.~S-M. and A.~P.
are grateful to Esteban Araya for his help in analyzing the data, to Francesco
Fontani and Asunci\'on Fuente for kindly providing complementary data, and also
to Rob Gutermuth for his help with Spitzer data. \'A.~S-M., A.~P. and R.~E. are
supported by the Spanish MICINN grant AYA2008-06189-C03 (co-funded with FEDER
funds). A.~P. is supported by a JAE-Doc CSIC fellowship co-funded with the
European Social Fund. This research has been partially funded by Spanish MICINN
under the ESP2007-65475-C02-02 and Consolider-CSD2006-00070 grants.  S.K. is
partially supported by DGAPA-UNAM grant IN101310. The National Radio Astronomy
Observatory is a facility of the National Science Foundation operated under
cooperative agreement by Associated Universities, Inc.
\end{small}




\end{document}